\documentclass[aps, pre, twocolumn, superscriptaddress, reprint]{revtex4-2}
\usepackage[T1]{fontenc}
\usepackage{times}
\usepackage{xfrac}
\usepackage{graphicx}
\usepackage{bm}

\usepackage{amsfonts, amsmath,amssymb, bm, graphicx}
\usepackage{siunitx}

\usepackage[english]{babel}
\usepackage{lipsum}

\newcommand{\figref}[2]{\hyperref[#1]{\ref{#1}(#2)}}
\usepackage{physics}
\usepackage{lipsum}
\usepackage{changes}

\usepackage[colorlinks=true,allcolors=blue]{hyperref}
\usepackage{footmisc}

\usepackage{upgreek}

\usepackage{soul}

\begin{document}

\makeatletter
\def\frontmatter@thefootnote{%
 \altaffilletter@sw{\@fnsymbol}{\@fnsymbol}{\csname c@\@mpfn\endcsname}%
}%

\makeatother
\title{Curvature-induced parity loss and hybridization of magnons:\\ Exploring the connection of flat and tubular magnetic shells}

\author{Felipe Brevis}
\affiliation{Helmholtz-Zentrum Dresden - Rossendorf, Institut f\"ur Ionenstrahlphysik und Materialforschung, D-01328 Dresden, Germany}
\affiliation{Departamento de Física, Universidad Técnica Federico Santa María, Avenida España 1680, Valparaíso, Chile}

\author{Pedro Landeros}
\affiliation{Departamento de Física, Universidad Técnica Federico Santa María, Avenida España 1680, Valparaíso, Chile}

\author{Jürgen Lindner}
\affiliation{Helmholtz-Zentrum Dresden - Rossendorf, Institut f\"ur Ionenstrahlphysik und Materialforschung, D-01328 Dresden, Germany}

\author{Attila Kákay}
\affiliation{Helmholtz-Zentrum Dresden - Rossendorf, Institut f\"ur Ionenstrahlphysik und Materialforschung, D-01328 Dresden, Germany}

\author{Lukas K\"orber}\email{lukas.koerber@ru.nl}
\affiliation{Helmholtz-Zentrum Dresden - Rossendorf, Institut f\"ur Ionenstrahlphysik und Materialforschung, D-01328 Dresden, Germany}
\affiliation{Radboud University, Institute of Molecules and Materials, Heyendaalseweg 135, 6525 AJ Nijmegen, The Netherlands}

\date{\today}

\begin{abstract}

This paper delves into the connection between flat and curvilinear magnetization dynamics. For this, we numerically study the evolution of the magnon spectrum of rectangular waveguides upon rolling its cross-section up to a full tube. Magnon spectra are calculated over a wide range of magnetization states using a finite-element dynamic-matrix method, which allows us to trace the evolution of the magnon frequencies and several critical magnetic fields with increasing curvature. By analyzing the parity of the higher-order magnon modes, we find a curvature-induced mode hetero-symmetry that originates from a chiral contribution to the exchange interaction and is related to the Berry phase of magnons in closed loops. Importantly, this curvature-induced parity loss has profound consequences for the linear coupling between different propagating magnons, allowing for hybridization between initially orthogonal modes. In this context, we demonstrate the integral role of edge modes in forming the magnon spectrum in full tubes. Our findings provide new theoretical insights into curvilinear magnetization dynamics and are relevant for interpreting and designing experiments in the field.

\end{abstract}

\maketitle

\section{Introduction}

Curvilinear systems in condensed matter exhibit a variety of emergent phenomena that are often not present in planar systems of the same material. Changes to topology, symmetry, or, in general, geometry that are induced by surface curvature can lead to many profound consequences. Examples include Landau levels in the electronic band structure of graphene \cite{niggeRoomTemperatureStraininduced2019}, geometric frustration in nematic crystals \cite{lopez-leonFrustratedNematicOrder2011}, correlated vortex tubes in superconductors \cite{fominTunableGenerationCorrelated2012}, or emergent anisotropies and chiral symmetry breaking in ferromagnets \cite{makarovMagnetismCurvedSurfaces2013,hertelCurvatureInducedMagnetochirality2013,makarovCurvilinearMicromagnetismFundamentals2022}, to name a few. Indeed, such geometrical effects are ubiquitous in condensed matter and can be found even in novel classes of materials, such as the recently discovered altermagnets, for which curvature-induced magnetization was predicted \cite{yershov2024curvature}. 

Emergent magneto-chirality and changes in the topology of ferromagnets can significantly alter the dynamics of their fundamental low-energy excitations, referred to as spin waves or magnons. Apart from their basic relevance for the dynamics of ferromagnets, these excitations have been proposed in various applications \cite{Flebus24} such as magnetic switching \cite{kammererFastSpinwavemediatedMagnetic2012}, neuromorphic computing \cite{papp_nanoscale_2021,korberPatternRecognitionReciprocal2023}, or quantum sensing and computing \cite{hetenyiLongdistanceCouplingSpin2022,doi:10.1126/sciadv.adi2042,10.1063/5.0157520}. While curvature-induced anisotropy can, for example, facilitate magnon propagation with high group velocities in the absence of external fields \cite{turcanSpinWavePropagation2021}, emergent chiral interactions -- both due to short-range or long-range interactions -- often result in nonreciprocal propagation accompanied by an asymmetric dispersion \cite{gaidideiMagnetizationNarrowRibbons2017,otaloraCurvatureInducedAsymmetricSpinWave2016,salazar-cardonaNonreciprocitySpinWaves2021} and altered linear coupling (hybridization) between different magnon modes \cite{gallardoHighSpinwaveAsymmetry2022,korberCurvilinearSpinwaveDynamics2022}. In specific situations, curvature-induced chiral symmetry breaking can also modify nonlinear magnon-magnon interactions \cite{korberCurvilinearSpinwaveDynamics2022}. 

 \begin{figure}
    \centering
    \includegraphics{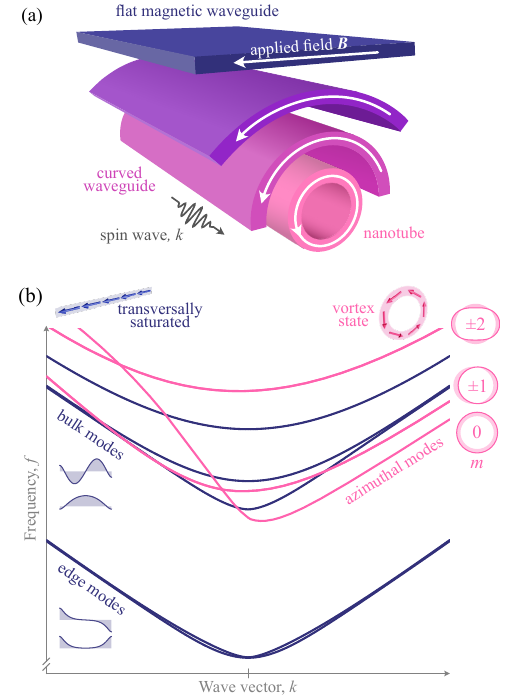}
    \caption{{(a) Schematic representation of the waveguides studied as the flat waveguide (blue) is rolled up to form a closed tube (pink). (b) Spin-wave dispersions for flat and tubular waveguides with transversally-saturated and vortex-state equilibrium magnetization, respectively, and some mode profiles.}}
    \label{fig:fig1}
\end{figure} 

This paper theoretically explores the continuous transition between the magnetization statics and dynamics in flat and curvilinear geometries, thereby uncovering an intimate connection between phenomena that emerge in both cases. Our study focuses on the propagation of magnons along waveguides with an initially rectangular cross-section that is gradually rolled up and curved to form a closed tubular system [seen in Fig.~\figref{fig:fig1}{a}]. The behavior and characterization of magnons for planar and tubular waveguides have been well studied but always viewed separately. Indeed, the magnon spectrum of flat rectangular waveguides and closed tubes is drastically different. These differences can depend on the arrangement of the equilibrium magnetization of the sample, upon which the magnons propagate. In the most straightforward scenario, this equilibrium can be controlled by externally applied transversal fields, as shown in Fig.~\figref{fig:fig1}{a}. The modes propagating along transversally-magnetized flat rectangular waveguides possess a fully symmetric dispersion $\omega(k)=\omega(-k)$, divided into a doublet of edge modes localized at the lateral boundaries of the cross-section, as well as an infinite set of discrete and nondegenerate bulk modes [see Fig.~\figref{fig:fig1}{b}, insets show line traces of the wave profiles along the transversal direction].
In contrast to this, the modes in closed nanotubes in the flux-closing vortex state exhibit a curvature-induced dispersion asymmetry [$\omega_m(k)\neq\omega_m(-k)$] and are characterized by their azimuthal mode index $m \in \mathbb{Z}$ in the azimuthal direction [see the insets in Fig.~\figref{fig:fig1}{b}]. Due to their nonreciprocal properties, these magnon modes in nanotubes can even show entirely unidirectional propagation in a certain frequency range \cite{korberCurvilinearSpinwaveDynamics2022,gallardoUnidirectionalChiralMagnonics2022}. While all modes in vortex-state tubes exhibit an asymmetric dispersion along the waveguide axis, modes with opposing azimuthal indices $\pm m$ are degenerate [$\omega_{m}(k)=\omega_{-m}(k)$]. As soon as the equilibrium magnetization in the tubular waveguide departs from the vortex state and acquires a nonzero component along the tube's axis, this degeneracy is lifted by a topological Berry phase that arises from emergent chiral interactions and is absent in the planar case \cite{hertelDomainWallInducedPhase2004,dugaevBerryPhaseMagnons2005,otaloraCurvatureInducedAsymmetricSpinWave2016,hillChiralMagnetismGeometric2021}.  

Despite the big qualitative difference between planar and tubular systems, a robust connection between the two cases can be established through inspection of the critical fields and their dependence on curvature, the behavior and symmetry of the modes for ferromagnetic resonance ($k=0$) and complete dispersion ($k \neq 0$). A finite-element matrix-dynamic method makes the numerical evaluation of these spectra possible. The particular magnetic system and numerical method used to calculate the magnon spectra for this study are described in detail in Sec.~\ref{sec:methods} of this paper. After that, in Sec.~\ref{sec:critical_fields}, we discuss the general curvature-dependence of different critical magnetic fields associated with the spectrum of the fundamental mode, which crucially determines the stability of the magnetic equilibrium. With that, we demonstrate how the resonance curve of a flat waveguide, upon rolling it up, smoothly approaches that of a closed tube. Following, in Sec.~\ref{sec:parityloss}, we turn to the higher-order modes and categorize them according to their parity. We predict a parity loss with increasing curvature and below the transversal saturation field. The consequential hetero-symmetry of the modes in this field regime connects to a curvature-induced chiral contribution to the exchange interaction, which allows us to directly relate it to the Berry phase of magnons in closed tubes. Above the saturation field, parity is well-preserved, which leads to the disappearance of odd-parity modes when the system's topology changes by closing the loop. The difference in parity behavior has substantial implications for hybridizing the propagating modes. Therefore, in Sec.~\ref{sec:dispersion}, we close by studying the dispersion of the propagating modes upon rolling up in these different field/parity regimes.
On the one hand, this shows the integral role of the edge modes of flat waveguides in forming the spectrum of full tubes by hybridizing with the bulk spectrum. On the other hand, we demonstrate how the parity loss mentioned above alters which modes hybridize depending on the strength of the external field. Our findings are summarized again in Sec.~\ref{sec:summary}.

By continuously bridging the gap between flat waveguides and nanotubes, we discuss the interplay of symmetry, curvature, and topological effects, providing new insights into curvilinear magnetization dynamics. Moreover, our predictions for the intermediate cases of open tubular segments apply to similar geometries, such as rolled-up tubes \cite{Bermude_Urena_2009,PhysRevLett.104.037205,bahlhorn2012rolled,PhysRevB.88.054402,streubel2014rolledup}, crescent-shaped nanorods \cite{golebiewskiSpinWaveSpectralAnalysis2023}, trenches, or corrugated surfaces \cite{turcanSpinWavePropagation2021}. These systems can be easier to manufacture than closed nanotubes, making our results relevant for designing experiments in curvilinear magnonics.

\section{Methodology}\label{sec:methods}

\subsection{Studied magnetic system}\label{sec:system}

\begin{figure}[h]
    \centering
    \includegraphics[width=8.6cm]{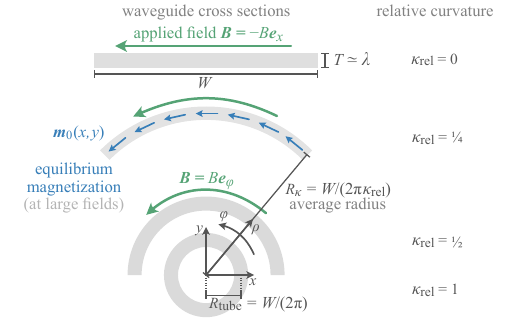}
    \caption{{Schematics showing the cross-sections of the waveguides studied according to the relative curvature with their respective applied external magnetic fields and associated equilibrium magnetization.}}
    \label{fig:schematics}
\end{figure}

We consider spin waves propagating along soft magnetic waveguides that are infinite (translationally invariant) along the $z$-axis as shown in Fig.~\ref{fig:fig1}, with the cross-sections depicted in Fig.~\ref{fig:schematics}. The rectangular waveguide has a width $W=\SI{160}{\nano\meter}$ and a thickness $T=\SI{10}{\nano\meter}$, which is in the order of the dipole-exchange length $\lambda=\sqrt{2A/\mu_0M_\mathrm{s}^2}$ of the respective material. The latter assumption is made for simplicity, such that we can assume all magnetization statics and dynamics to be homogeneous along the out-of-plane direction of the waveguide. For our calculations below, we assume typical material parameters of Ni$_{80}$Fe$_{20}$ (permalloy): saturation magnetization $\mu_0M_{\rm{s}} = \SI{1}{\tesla}$, exchange stiffness constant $A=\SI{13}{\pico\joule/\meter}$, and gyromagnetic ratio $\gamma = 176.086$ rad GHz/T, resulting in a dipole-exchange length $\lambda\approx \SI{5.72}{\nano\meter}$.

Upon rolling up the waveguide, the central arc length and thickness of the bent cross-section are kept constant, ultimately resulting in a full tube with central circumference $W$, average radius $R_\circ=W/2\pi$ and shell thickness $T$ [see Fig. \ref{fig:schematics}]. The transition between the two geometries is parameterized by the relative curvature $0\leq \kappa_\mathrm{rel} \leq 1$ defined as $\kappa_\mathrm{rel}=R_\kappa/R_\circ$, the curvature relative to that of the closed tube. This definition gives rise to the average curvature radius of each tubular segment $R_{\kappa}=W/(2\pi\kappa_\mathrm{rel}) = R_\circ/\kappa_\mathrm{rel}$ (see Fig.~\ref{fig:schematics}). Obviously, the limiting cases $\kappa_\mathrm{rel}=0$ and $\kappa_\mathrm{rel}=1$ describe the flat rectangular and fully tubular cases, respectively.

The specific magnetic equilibrium state $\bm{M}_0(x,y)$ within the cross-sections of the curved waveguides has crucial implications on the spin-wave spectrum. Therefore, an external magnetic field $\bm{B}$ in the azimuthal ($\varphi$) direction is applied to control the equilibrium. Such a field could be experimentally achieved by placing the waveguides on/around a conducting wire. In the limit of $R_\kappa \rightarrow \infty$, for the rectangular waveguide, the external homogeneous field points along the transversal direction of the waveguide (negative $x$-direction). This, in return, can be experimentally achieved by placing the waveguide on a flat strip-line antenna. Depending on the strength of this static field, the unitary equilibrium magnetization $\bm{m}_0(x,y)=\bm{M}_0(x,y)/M_\mathrm{s}$ can either be axially saturated ($z$-direction), transversally saturated ($\varphi$ or $x$-direction), or in some intermediate state. This field-dependence will be discussed in more detail in Sec.~\ref{sec:critical_fields}.


\subsection{Micromagnetic modeling}

\begin{figure}[h]
    \centering
    \includegraphics[width=8.6cm]{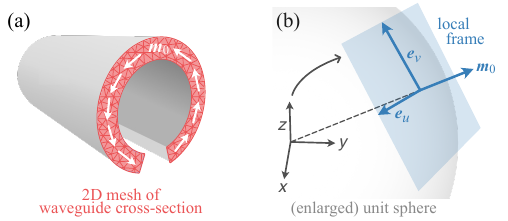}
    \caption{(a) Schematic two-dimensional mesh used to model the different waveguides using a finite-element dynamic-matrix method. (b) Frame of reference locally orthogonal to the equilibrium direction $\bm{m}_0$ at each point of the cross-section (adapted from Ref.~\citenum{korberSpinWavesCurved2023}).}
    \label{fig:dynmat}
\end{figure}

To study the linear magnon spectra in the curved waveguides, we employ a finite-element dynamic-matrix method for propagating spin waves, developed in Ref.~\citenum{korberFiniteelementDynamicmatrixApproach2021a} and implemented in the micromagnetic modeling package \textsc{TetraX} \cite{korberTetraXFiniteElementMicromagneticModeling2022}. This method obtains the angular frequency $\omega_\nu$ and the unitless complex-valued spatial mode profile $\bm{m}_\nu(\bm{r})$ of the $\nu$th magnon eigenmode by numerically solving the linearized Landau-Lifshitz equation
\begin{equation}\label{eq:linllg}
    \omega_\nu \bm{m}_\nu = i\gamma\, \bm{m}_0 \times \vu{\Omega}\cdot \bm{m}_\nu \quad\text{with}\quad\bm{m}_\nu\perp\bm{m}_0
\end{equation}
in the vicinity of a stable equilibrium magnetization $\bm{m}_0$. We calculate the equilibrium by minimizing the total magnetic energy for each value of the external field $\bm{B}$. The operator $\vu{\Omega}$, the Hessian of this energy in $\bm{m}_0$, can be written as
\begin{equation}
    \vu{\Omega}=  \mu_0M_\mathrm{s}\vu{N} + B_0\vu{I}
\end{equation}
with $ \vu{N}$ being the magnetic tensor, a unitless Hermitian operator that describes the magnetic self-interactions and yields the internal magnetic field due to these interactions as $\bm{B}_\mathrm{int}=-\mu_0M_\mathrm{s}\vu{N}\cdot\bm{m}$. Furthermore, $B_0 = \bm{m}_0\cdot[ \bm{B}_\mathrm{int}(\bm{m}_0) + \bm{B}]$ is the projection of the total effective equilibrium field, composed of the internal field and the externally applied field, onto the equilibrium magnetization. For our present study, we only consider the exchange ($\vu{N}_\mathrm{x}$) and dipolar ($\vu{N}_\mathrm{d}$) magnetic self-interactions, allowing us to write the magnetic tensor as  
\begin{equation}
    \vu{N} = \vu{N}_\mathrm{x} + \vu{N}_\mathrm{d} = - \lambda^2\bm{\nabla}^2 + \nabla\phi[\,.\,]
\end{equation}
with the dipole-exchange length $\lambda$ defined before and $\phi[\bm{m}]$ being the magnetostatic potential generated by the volume and surface divergencies of the magnetization $\bm{m}$. In \textsc{TetraX}, this dipolar potential is calculated using the hybrid finite-element/boundary-element method for propagating waves developed in Ref.~\citenum{korberFiniteelementDynamicmatrixApproach2021a}. Assuming translational symmetry of the magnetic waveguides along the $z$-direction, the spatial profiles can be written as
\begin{equation}
    \bm{m}_\nu (\bm{r}) = \bm{\eta}_\nu(\bm{\rho},k)e^{ikz} 
\end{equation}
 with $\bm{\eta}_\nu$ being the lateral profile of the $\nu$th mode at wave number $k$, which depends on the coordinates $\bm{\rho}=(x,y)$ in the cross-section of the waveguide [see Fig.~\figref{fig:dynmat}{a}]. This ansatz allows us to transform the linearized equation~\eqref{eq:linllg} to a single cross-section and solve it on this domain for each $k$. As a result, one numerically obtains the exact dispersion $\omega_\nu(k)$ of the different propagating modes along the waveguides. Each lateral profile also needs to satisfy the constraint $\bm{\eta}_\nu\perp\bm{m}_0$, which is achieved by additionally projecting Eq.~\eqref{eq:linllg} into the frame of reference ${\bm{e}_u, \bm{e}_v}$ that is locally orthogonal to the equilibrium direction $\bm{m}_0$ [see Fig.~\figref{fig:dynmat}{b}]. Further details on the eigenvalue problem and its numerical implementation in finite elements can be found in Refs.~\citenum{korberFiniteelementDynamicmatrixApproach2021a,korberFiniteelementDynamicmatrixApproach2022}. For accuracy, the average edge length in the finite-element discretization is set to \SI{2}{\nano\meter}, lower than the dipole-exchange length $\lambda \approx \SI{5.72}{\nano\meter}$ corresponding to the material parameters described in Sec.~\ref{sec:system}. 

\section{Results and discussion}

\subsection{Curvature-dependence of critical fields}\label{sec:critical_fields}


Before diving into the peculiarities of the full magnon spectrum, we start by discussing the evolution of the fundamental mode (ferromagnetic resonance mode), which oscillates with a homogeneous phase across the waveguide. This will allow us to discuss the basic dependence of the magnetization on the applied field and curvature. 

\begin{figure}[h!]
    \centering
    \includegraphics{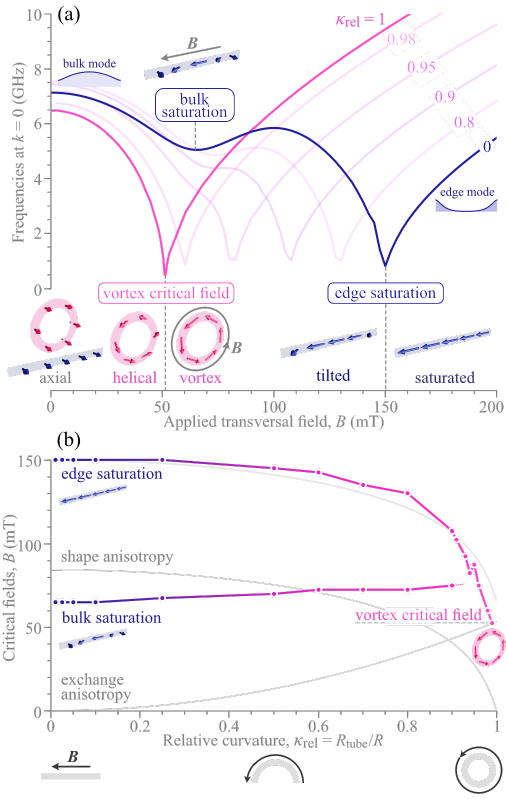}
    \caption{(a) Dependence of the fundamental mode frequency on the applied transversal external field for different values of the relative curvature $\kappa_\mathrm{rel}$. Thickness and arc length of the cross-sections are fixed at $T=\SI{10}{\nano\meter}$ and $W=\SI{160}{\nano\meter}$. Insets for the flat case show the transformation of this mode from a bulk into an edge mode, undergoing two characteristic frequency minima. Insets with arrows sketch the equilibrium magnetization in different regimes. (b) Evolution of the critical fields, bulk- and edge-saturation field on the relative curvature $\kappa_\mathrm{rel}$, during the transition between a flat waveguide to a tube.}
    \label{fig:minimavskappa}
\end{figure}

To this end, figure \figref{fig:minimavskappa}{a} shows the dependence of the fundamental-mode frequency on the applied transversal field for different values of the relative curvature $\kappa_\mathrm{rel}$. The thickness and width of the cross-section are set to $T=\SI{10}{\nano\meter}$ and $W=\SI{160}{\nano\meter}$, respectively. In the planar case, $\kappa_\mathrm{rel}=0$, and at zero field, the bulk mode exhibits a partially-pinned sinusoidal profile across the width direction of the waveguide and, with increasing field, transforms into a symmetric edge mode, localized at the lateral boundaries of the waveguide cross-section. This change in the spatial mode profile is inset as line traces in Figure \figref{fig:minimavskappa}{a}. In the intermediate field range, the frequency of the fundamental mode undergoes two characteristic minima at critical fields denoted as the \textit{bulk-saturation} and the \textit{edge-saturation} field, respectively. At the bulk-saturation field, mainly determined by the competition between Zeeman, exchange and dipolar interactions, the magnetization in the middle of the waveguide rotates towards the external field. As the field is further increased, the resonance frequency of the lowest mode increases. This increase can be attributed to an increase of the exchange interaction contribution to the mode energy, due to the gradual confinement of the dynamical region between the bulk-saturated region and the waveguide edges \cite{iurchuk2024anatomylocalizededgemodes}. At about \SI{100}{\milli\tesla} the exchange contribution reaches its maximum and the further increasing demagnetizing field due to the continuous rotation of the magnetization at the edges towards the external field direction, will decrease the mode frequency.  At the second minimum, the edge-saturation field, the edge region becomes fully saturated. Due to the zero internal fields at this field, the frequency goes down to zero and the mode becomes soft (a Goldstone mode \cite{PhysRev.127.965}). The curves in Fig.~\figref{fig:minimavskappa}{a} do not go precisely to zero due to the finite step size when changing the external field. Above this saturation, the edge mode behaves as a Kittel-like mode. Therefore, the further field increase does not impact the mode localization but only leads to an almost frequency increase almost linear in the field. Note, at the bulk-saturation field, even for very wide stripes, the frequency minima will never reach zero due to the excess exchange energy originating from the domain wall-like region of the magnetization between the bulk and the edges.

Upon rolling up the planar waveguide, the relative curvature $\kappa_\mathrm{rel}$ increases, and these two critical fields are altered in distinct ways. While the bulk-saturation field increases weakly, the edge-saturation field strongly decreases. Once the tube is closed at $\kappa_\mathrm{rel}=1$, the fundamental-mode frequency exhibits only a single minimum at the vortex critical field [see Fig.~\figref{fig:minimavskappa}{a}] which separates the fully transversally-saturated vortex state from the global helical state a lower fields. In the global helical state (which corresponds to a homogeneously tilted magnetization when unrolling the tube) the magnetization partially curls the $\varphi$-direction while retaining a $z$-component \cite{salazar-cardonaNonreciprocitySpinWaves2021}.

The intermediate transition of the critical fields from flat waveguide to closed tube is reported for more curvature values in Figure \figref{fig:minimavskappa}{b}, which shows the critical-field values as a function of curvature. For $\kappa_\mathrm{rel}>0.9$, the edge-saturation minimum completely dominates the spectrum, making extraction of the bulk minimum impossible. However, being determined mainly by the competition between dipole and exchange interactions, it remains almost constant across the curvature range.

For a better understanding of the evolution of the edge-saturation field, it is helpful to consider two critical fields that influence this transition, namely the approximate demagnetization field $B_\mathrm{shape}$ resulting in a nonlocal shape anisotropy and the exchange field $B_\text{x-an}$ resulting in a local curvature-induced anisotropy. The shape anisotropy field is produced by magnetic surface charges at the lateral edges of the waveguides. It counteracts the orientation of the magnetization perpendicular to the surface, and therefore, saturation in the transversal direction (either $x$ or $\varphi$, depending on whether $\kappa_\mathrm{rel}=0$ or not). In the macro-spin approximation, we assume that the magnetization is homogeneous within the cylindrical frame of reference $\bm{m}_0 = m_\varphi\bm{e}_\varphi + m_z\bm{e}_z$ where the components $m_\varphi$ and $m_z$ depend on the external field $\bm{B}$ but not on the coordinates $\bm{\rho}=(x,y)$. For this case, the shape anisotropy field is given by 
\begin{equation}
    B_\mathrm{shape} =  -\mu_0 M_\mathrm{s} \mathcal{N}_{\varphi\varphi} 
\end{equation}
with
\begin{equation}
    \mathcal{N}_{\varphi\varphi} 
    = \expval{\bm{e}_\varphi\cdot\vu{N}_\mathrm{d}\cdot\bm{e}_\varphi}
\end{equation}
being the transversal demagnetizing factor in the macro-spin approximation, a matrix element of the dipolar tensor $\vu{N}$. The spatial average $\expval{...}$ is taken within the cross-section of each waveguide. For zero curvature $\kappa_\mathrm{rel}=0$, this demagnetizing factor was found analytically by Aharoni \cite{aharoniDemagnetizingFactorsRectangular1998}. Here, for any $\kappa_\mathrm{rel}$, we calculate it numerically with \textsc{TetraX}, which implements a numerical representation of the dipolar field using a plane-wave hybrid finite-element/boundary-element method \cite{korberFiniteelementDynamicmatrixApproach2021a}. Notably, for $\kappa_\mathrm{rel}=0$, the numerical result from \textsc{TetraX} agrees perfectly with the analytical formulas in Ref.~\cite{aharoniDemagnetizingFactorsRectangular1998}. We see in Fig.~\figref{fig:minimavskappa}{b} that the edge-saturation field mostly follows the trend of the shape anisotropy but with an almost constant offset. This offset is because, when applying a transversal field, the magnetization does not evolve according to the macro-spin approximation. Instead, first only the middle of the waveguide rotates into the direction of the external field, while a wide region close to the edges is still magnetized along the strip, forming a kind of a 90-degree Néel wall between the middle region and the edge region. With increasing field, this domain wall is pushed towards the edges and confined as it reaches the edges. This equilibrium state has an elevated exchange contribution compared to a macrospin approximation. The edge-saturation field, however, approaches the macro-spin field for waveguides with aspect ratio $W/T\gg 1$ or, in the rectangular case, for exchange-dominated geometries where $T$ and $W$ are in the order of the dipole-exchange length of the material or smaller. This fact is used, for example, in calculating the dipolar field using the demagnetizing factors of parallelepipeds in finite-difference methods \cite{henryPropagatingSpinwaveNormal2016}.

Although the shape-anisotropy field in the macro-spin approximation describes the evolution of the edge-saturation field for our waveguides only poorly up to a constant offset, it provides the correct intuition: As the relative curvature increases and the rectangular waveguide is rolled up to a tube, it becomes easier to close the flux produced by stray field lines exciting and entering the lateral edges of the waveguide, until, at $\kappa_\mathrm{rel}=1$, the flux can be closed entirely in the vortex state $\bm{m}_0=\bm{e}_\varphi$. This is why the shape-anisotropy field vanishes as $\kappa_\mathrm{rel}$ approaches unity. In contrast, however, the edge-saturation field does not vanish but instead approaches a constant value known as the critical vortex field or the exchange field of the vortex. This field acts as an effective local anisotropy that penalizes curling the magnetization along the $\varphi$-direction. Curvature-induced anisotropy is well-understood in curvilinear magnetism and can be described by expressing the exchange energy, or the exchange operator, in the frame of reference that is locally attached to the curvilinear shell. Then, the exchange operator $\vu{N}_\text{x} = -\lambda^2\bm{\nabla}^2$ separates into three contributions,
\begin{equation}
    \vu{N}_\text{x} = \vu{N}_\text{x-cov} + \vu{N}_\text{x-an} + \vu{N}_\text{x-ch}
\end{equation}
where $\vu{N}_\text{x-cov}=-\lambda^2 \vu{I}\nabla^2$ is due to the variation of the magnetization in the curvilinear frame of reference (in terms of the covariant derivative), $\vu{N}_\text{x-an}$ is the emergent anisotropy and $\vu{N}_\text{x-ch}$ represents emergent chiral interactions, known as curvature-induced Dzyaloshinskii-Moriya interactions (DMI) of exchange type. Dipolar coupling may also lead to DMI-like contributions which we will discuss in the later part of this paper. For the moment, only the anisotropy-like contribution $\vu{N}_\text{x-an}$ to the exchange interaction is important. In a cylindrical frame of frame of reference, it reads
\begin{equation}
    \vu{N}_\text{x-an}  =  {\mqty(\frac{\lambda^2}{\rho^2} & 0 & 0 \\
    0 & \frac{\lambda^2}{\rho^2} & 0 \\
    0 & 0 & 0)}.
\end{equation}
Then, the averaged field $B_\text{x-an} = -\mu_0M_\mathrm{s}\langle \bm{e}_\varphi\vu{N}_\text{x-an}  \bm{e}_\varphi\rangle$ associated with entirely aligning the magnetization along the $\varphi$-direction is given by
\begin{equation}
    B_\text{x-an}\  = -{\mu_0 M_\mathrm{s}\lambda^2}{\expval{\rho^{-2}}_\rho} \approx -\frac{2A}{M_s R^2_\circ}\ \kappa_\mathrm{rel}^2 
\end{equation} 
where $\expval{\rho^{-2}}_\rho$ is the squared inverse radial coordinate, averaged along the radial direction. For curvature-radii $R\gg T$ of the tubular segments, it can be approximated as $R^2 = R_\circ^2/\kappa_\mathrm{rel}^2$ with $R_\circ$ being the average radius of the full tube (for more exact expressions of the anisotropy field in tubes, see for example, Refs.~\citenum{otaloraCurvatureInducedAsymmetricSpinWave2016, otaloraAsymmetricSpinwaveDispersion2017, korberCurvilinearSpinwaveDynamics2022}). The curvature-induced anisotropy and the chiral interaction are approximately quadratic in the relative curvature  $\kappa_\mathrm{rel}$. However, the latter can be neglected when considering the frequency dependence of the fundamental mode (whose spatial profile is almost entirely homogeneous). Therefore, as also found by inspecting the fundamental-mode frequency evolution in the tubular waveguides in Fig.~\figref{fig:minimavskappa}{b}, it uniquely determines the saturation field as the tube is closed,  and $\kappa_\mathrm{rel}=1$. We will see, however, in the next section that the emergent chiral interaction plays a crucial role in the dynamics of the higher-order standing modes.

\subsection{Hetero-symmetry of higher-order standing modes}\label{sec:parityloss}

In the following, we go beyond the evolution of the fundamental mode only and consider modes that can be inhomogeneous with the cross-section of the waveguides (all of them will still have $k=0$ wave number along the waveguide axes). We shall see that the emergent chiral interaction discussed in the previous section will modify the symmetries of the modes when departing from the planar case $\kappa_\mathrm{rel}=0$. The symmetries of the different modes, in return, determine their evolution when changing $\kappa_\mathrm{rel}$ as well as their inter-mode hybridization and  susceptibility to external microwave fields.

\begin{figure*}
    \centering
    \includegraphics[width=17.5cm]{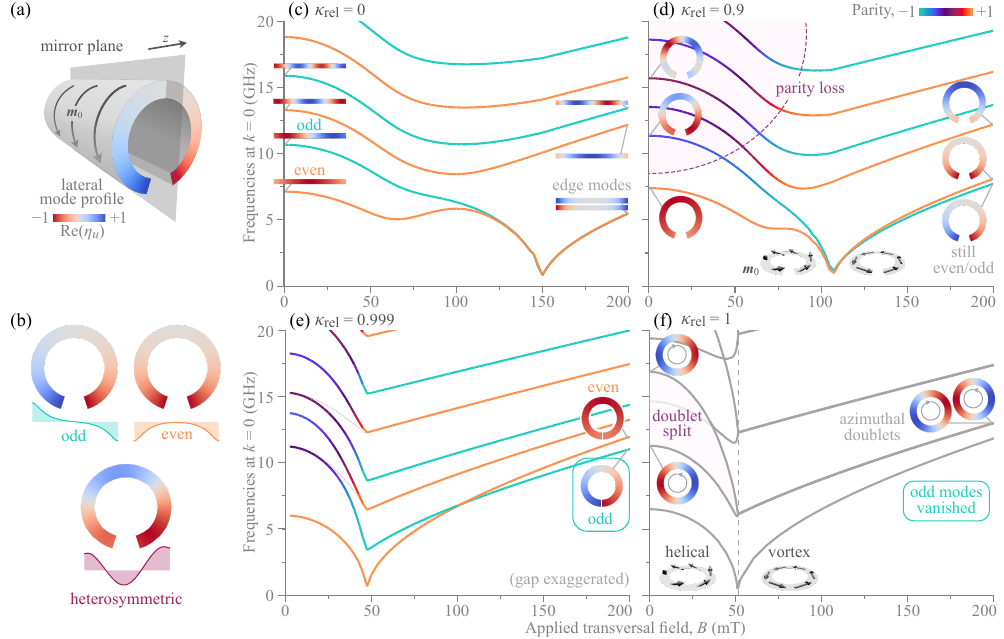}
    \caption{{Parity and ferromagnetic resonance for different values of $\kappa_{\textrm{rel}}$. The parity calculation compares the magnetization profile in the local coordinate system for each mode considering a mirror plane, as shown in (a). Based on this comparison, the modes can be odd ($P=-1$), even ($P=+1$) or hetero-symmetric (a value in between), as shown in (b) with their respective mode profiles colored according to parity. (c-f) Ferromagnetic resonances for waveguides with $\kappa_{\textrm{rel}} =$ 0, 0.9, 0.999 and 1. Modes are colored according to parity except for $\kappa_{\textrm{rel}}=1$, where parity is meaningless. Insets indicate the mode profiles for different applied fields and equilibrium magnetization are also depicted.}} 
    \label{fig:fmr_kappa_parity}
\end{figure*}


To establish a symmetry categorization of the different magnon modes, we realize that there is one mirror plane, the $yz$ plane, which remains intact when rolling up the planar waveguide to a full tube [see Fig.~\figref{fig:fmr_kappa_parity}{a}]. Note that this mirror plane is, strictly speaking, only a symmetry of the geometry itself, not the full magnetic system. Depending on the magnetization state $\bm{m}_0(B)$ in the cross-section, the plane can be part of its magnetic point group as a simple mirror plane or, in other cases, only when combined with an additional time reversal $t\mapsto -t$. The latter situation appears hand in hand with the emergence of curvature-induced chirality. 

To quantify the mode symmetry we consider that parity of the different modes with respect to this mirror plane. Therefore, similar to Ref.~\citenum{trevillian_formation_2024}, we consider the expectation value of the parity operator $\hat{P}$ that takes $x\mapsto -x$ as
\begin{equation}\label{eq:parity}
    P_\nu := \langle \Tilde{\bm{\eta}}_\nu \vert \hat{P}\vert\Tilde{\bm{\eta}}_\nu\rangle = \int\limits_\mathrm{A} \mathrm{d}x\,\mathrm{d}y\ \Tilde{\bm{\eta}}^*_\nu(x,y)\cdot \Tilde{\bm{\eta}}_\nu(-x,y)
\end{equation}
where $\Tilde{\bm{\eta}}_\nu$ is the complex-valued lateral mode profile (see Sec.~\ref{sec:methods}) of the $\nu$th mode taken in the frame of reference locally orthogonal to the equilibrium magnetization $\bm{m}_0$. The profiles are normalized such that $\braket{\Tilde{\bm{\eta}}_\nu}{\Tilde{\bm{\eta}}_\nu}=1$. The parity of a mode according to Eq.~\eqref{eq:parity} can take values $-1 \leq P_\nu \leq 1$ while $P_\nu=\pm 1$ are taken by eigenvectors of the parity operator $\hat{P}$, that is, by modes that are either even ($+1$) or odd ($-1$) with respect to the mirror plane shown in Fig.~\figref{fig:fmr_kappa_parity}{a}. We will soon see that not all modes have definite parity, $P_\nu \neq \pm 1$, but instead take on intermediate values and are therefore hetero-symmetric. Examples for even, odd and hetero-symmetric modes are shown in Fig.~\figref{fig:fmr_kappa_parity}{b} as the real part of one component of the lateral mode profile ($\Re \Tilde{\eta}_{u}$) in the frame of reference $\{\bm{e}_u,\bm{e}_v\}$ locally orthogonal to $\bm{m}_0$. Line traces along the azimuthal direction are set below to highlight their symmetry. 

Starting with the planar case $\kappa_\mathrm{rel}=0$, in Fig.~\figref{fig:fmr_kappa_parity}{c}, we report the field-dependence of the lowest six modes with $k=0$ wave-vector along the waveguide axis. The lines are colored according to their parity and insets show the lateral profiles at specific values of the external field. Importantly, the parity of all modes is well-defined and preserved for all external fields. At $B=0$, in the axially-magnetized state, the higher-order modes above the fundamental mode are standing waves across the transversal direction of the waveguide, with an increasing number of nodal lines and alternating parity. Indeed, these profiles can be approximated well with partially pinned sinusoidals and boundary conditions determined by the aspect ratio of the cross-section \cite{roussigneExperimentalTheoreticalStudy2001, guslienkoEffectiveDipolarBoundary2002a}. As the field increases, these modes undergo a frequency minimum and smoothly transform into the spectrum of the transversally magnetized waveguide, with their mode profiles shown for $B=\SI{200}{\milli\tesla}$. Importantly, during this transition, the odd second-order mode joins the even fundamental mode to form a pair of degenerate edge modes for fields above saturation (about \SI{150}{\milli\tesla}). One can clearly see how the mode profiles have been transformed and the spectrum has been divided between modes localized to the edge and to the bulk. Interestingly, the lowest-order bulk mode (third mode in the spectrum) is not homogeneous with the cross-section but maintains the two nodal lines it already had at $B=0$.

Departing from the planar case and increasing the relative curvature to $\kappa_\mathrm{rel}=0.9$ in Fig.~\figref{fig:fmr_kappa_parity}{d}, the  degeneracy between the even and odd edge modes is lifted. Moreover, we observe the shifting of the minima in the spectrum of the fundamental mode as discussed in Sec.~\ref{sec:critical_fields}. Above the decreased saturation field of approximately \SI{110}{\milli\tesla}, the spectrum remains largely unchanged, with all modes pertaining their parity and, with the exception of the two edge modes, also their frequencies (although the curvature is already drastically changed). However, below \SI{110}{\milli\tesla}, we observe a parity loss of the higher-order modes, as indicated by the blue/purple coloring of their spectral lines and verified by the insets showing the lateral profiles at zero field. Clearly, these modes have become hetero-symmetric due to the curvature. This parity loss and associated hetero-symmetry of the higher-order modes can be understood as a consequence of the curvature-induced chiral contribution $\vu{N}_{\text{x-ch}}$ to the exchange interaction, that emerges in the curvilinear frame of reference. In the cylindrical frame, it is written as 
\begin{equation}
    \vu{N}_{\text{x-ch}} ={\mqty(0& \frac{2\lambda^2}{\rho^2}\pdv{}{\varphi} & 0 \\
   - \frac{2\lambda^2}{\rho^2}\pdv{}{\varphi} & 0 & 0 \\
    0 & 0 & 0)} 
\end{equation} 
This curvature-induced DMI potentially breaks the inversion symmetry in the azimuthal $\varphi$-direction ($+\varphi$ and $-\varphi$ are not equivalent) and, much like the curvature-induced anisotropy, is approximately quadratic in the relative curvature. Its origin can be traced back to the fact that the local frame is parallel-transported (or revolved) along the azimuthal direction of the curvilinear shell, which results in a spin connection that, in our case, couples the $\rho$ and the $\varphi$ components of the magnetization. Loosely speaking, going a small step along the $\varphi$-direction slightly rotates one direction into the other, which can result in a Berry phase of the precessing magnetization vector  \cite{dugaevBerryPhaseMagnons2005,hillChiralMagnetismGeometric2021} that can break the symmetry between modes counter-propagating along the $\varphi$-direction by a contribution proportional to $\lambda^2\expval{\rho^{-2}}_\rho\sim \kappa_\mathrm{rel}^2$.  Here, even when the loop is not yet closed and we are dealing with standing waves, we observe its effect as a mode parity loss. We remind that, lacking an exact analytical theory for the spin-wave spectrum in open tube segments, the discussion of the numerically obtained spectra is qualitative. However, the hetero-symmetry of the higher-order modes can be understood in the following way: With the higher-order modes being standing waves, they can be composed of two counter-propagating modes along the azimuthal direction. With the symmetry between $+\varphi$ and $-\varphi$-direction being broken, counter-propagating solutions at the same frequency will exhibit different wavelengths. Therefore, their superposition, in general, cannot form a symmetric standing wave. Clearly, the fundamental mode is completely unaffected by this, as it has homogeneous phase along the azimuthal direction cannot be composed of two counter-propagating modes. As the curvature-induced DMI only couples the $\rho$ and $\varphi$ component of the mode profiles $\bm{\eta}_\nu\perp\bm{m}_0$, it participates in the dispersions only as long as $\bm{m}_0$ has a nonzero $z$ component and is completely ineffective when the waveguide is saturated along $\varphi$ above \SI{110}{\milli\tesla} [see the insets in Fig.~\figref{fig:fmr_kappa_parity}{d} sketching the equilibrium states in the two field regimes]. Hence, in the $\varphi$-saturated state, the parity of all modes is preserved. This is fully consistent with the fact that, in this case, the $xz$ plane is a true mirror plane of the equilibrium state. That is, mirroring the state with respect to this plane superimposes $\bm{m}_0$ onto itself. However, this mirror symmetry is broken for any finite $z$ component of the equilibrium, as $\bm{m}_0$ is transformed into a state with inverted $z$ component. For the fully axially-polarized state $\bm{m}_0=\bm{e}_z$, however, the $xy$ plane represents a m' plane, that is, a plane that mirrors if combined with time reversal. In any case, only a finite $z$ component allows for chiral symmetry breaking along the $\varphi$-direction.

Almost closing the cross-section to a loop by increasing the relative curvature to $\kappa_\mathrm{rel}=0.999$ further exaggerates the qualitative differences between the two field regimes [see Fig.~\figref{fig:fmr_kappa_parity}{e}]. Notice how the bulk-saturation minimum has been eaten up by the edge-saturation dip in the frequency curve of the fundamental mode. Moreover, minima in the frequency of all modes are not almost perfectly aligned at the same field position. Before closing the loop to a full tube, it is important to appreciate how the modes still possess well-defined parity at large fields. The insets in Fig.~\figref{fig:fmr_kappa_parity}{e} show the  profiles of the two lowest even and odd modes, respectively. To make it more visible, the gap in the cross-section has been visually exaggerated. At these large fields, right before closing the loop to a full tube, even modes correspond to laterally standing waves with an integer number of periods $\abs{m}=0,1,2$, and so forth, along the azimuthal direction, whereas odd modes exhibit a half-integer period $\abs{m}=\sfrac{1}{2},\sfrac{3}{2},\sfrac{5}{2}$ and so forth. It is clear that, upon closing the loop to a full tube ($\kappa_\mathrm{rel}=1$), these modes would lead to discontinuities in the dynamic magnetization. Therefore, they vanish and are no longer part of the spectrum at large fields, seen in Fig.~\figref{fig:fmr_kappa_parity}{f}. Instead, the formerly even modes with full periods along the azimuthal direction have now become doublets, representing the fact that counter-propagating modes in the azimuthal direction $m=\pm 1, \pm 2, ...$ are degenerate in the $\varphi$-saturated (vortex) state. In Fig.~\figref{fig:fmr_kappa_parity}{f}, the spectrum is no longer colored, since mode parity is no longer well defined when cylindrical symmetry is fulfilled. A closed tube possesses infinite mirror planes (denoted by any fixed $0 \leq \varphi_0 < \pi$). For any mode with a given azimuthal index $m$, we can choose a mirror plane at $\varphi_0$ for which this mode is odd and, at the same time, another one at $\varphi_0 + m\pi/2$ for which the mode is even. Therefore, modes no longer have a definite parity in the closed tube.

In the field regime below the critical vortex field (here, \SI{51.1}{\milli\tesla}), of course, the equilibrium is still partly oriented in the $z$-direction and forms a global helical state [see the inset in Fig.~\figref{fig:fmr_kappa_parity}{f}]. In this regime, the newly formed azimuthal doublets are not degenerate but, as mentioned before, split by exactly the same contribution in the exchange interaction, namely the chiral contribution $\vu{N}_\text{a-ch}$ that leads to the parity loss of the higher-order modes before closing the loop ($\kappa_\mathrm{rel}<1$). For a closed tube, this splitting is well-known and corresponds to a nontrivial version of the famous Aharonov-Bohm effect \cite{aharonovSignificanceElectromagneticPotentials1959} that can also be found in tubes with polygonal cross-section \cite{korberModeSplittingSpin2022} or confined magnetic rings \cite{uzunovaNontrivialAharonovBohmEffect2023}. However, in contrast to flat disks, in infinite tubes and at $k=0$ (for axially homogeneous modes), only the exchange interaction contributes to this split \cite{salazar-cardonaNonreciprocitySpinWaves2021}. Indeed, in the axial state, the Berry phase that results from the chiral interaction $\vu{N}_\text{x-ch}$ is exactly $2\pi$ and, therefore, shifts the periods of the modes $m \mapsto m \pm 1$ (the sign depends on the polarity of the axial state). In the exchange-dominated approximation, and in terms of these new indices, the azimuthal modes then become degenerate in the axial state. For a more elaborate discussion on this Berry phase in tubes see, for example, Sec.~9.2 in Ref.~\citenum{korberSpinWavesCurved2023}.

After having reached the full tube, the reader is invited to retrace again the transition of the spectrum from $\kappa_\mathrm{rel}=0$ in Fig.~\figref{fig:fmr_kappa_parity}{c} to $\kappa_\mathrm{rel}=1$ in Fig.~\figref{fig:fmr_kappa_parity}{f}, which neatly shows how the parity loss, or curvature-induced hetero-symmetry, and the doublet split due to an Aharonov-Bohm flux are connected, the former being a "standing-wave version" of the latter. In some sense, the mode hetero-symmetry is more ubiquitous as it does not rely on the continuity of the waveguide cross-section but solely on the local curvature. We expect analogous observations in similar geometries with curved cross-sections. Indeed, a viable experimental platform to validate these predictions could be crescent-shaped nanorods \cite{golebiewskiSpinWaveSpectralAnalysis2023}. It is important to note that the parity loss observed here is fundamentally distinct from the hetero-symmetry induced by dynamic dipolar fields of the modes propagating with $\bm{k}\perp\bm{m}_0$ (Damon-Eshbach waves) in magnetic thin films or disks \cite{henryPropagatingSpinwaveNormal2016,PhysRevLett.122.117202}. In these systems, the mirror symmetry of the modal profiles along the normal direction of the magnetic layer is broken by the interplay of magnetic volume and surface pseudo-charges, ultimately resulting in dipole-dipole hybridization between different branches of the dispersion. This effect appears strictly only for modes with $k \neq 0$ while the curvature-induced parity loss observed here also appears for the modes at $k=0$, and, as we will see in the following, even across the entire wave-vector range. Indeed, our situation is more similar to the confined modes in systems with intrinsic Dzyaloshinskii-Moriya interaction \cite{PhysRevB.99.214429}.

\begin{figure*}
    \centering
    \includegraphics{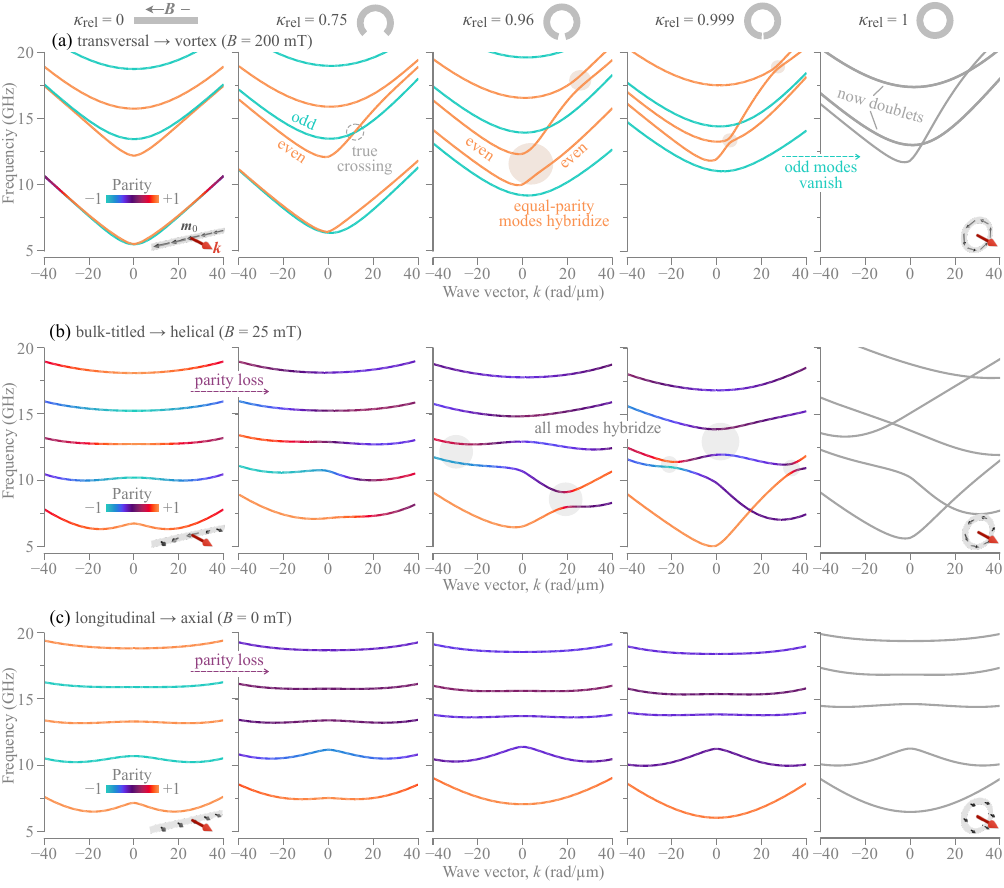}
    \caption{{Spin-wave dispersions for three different equilibrium magnetization regimes depending on the applied fields and for different values of the relative curvature ($\kappa_{\textrm{rel}} = 0,\,0.75,\, 0.96,\, 0.999,\,1$): (a) from transverse to vortex state at $B =$ 200 mT, (b) from bulk-tilted to global helical state at $B = $25 mT and (c) from longitudinal to axial state at $B = 0$. Modes are colored according to parity and insets depict the equilibrium magnetization. Grey-shaded circles indicate anti-crossings between different modes.}  }
    \label{fig:dispersion_evolution}
\end{figure*}

\subsection{Hybridization and dispersion of propagating modes}\label{sec:dispersion}

Up to now, we have seen that the spectrum of the magnon modes with $k=0$ along flat rectangular and tubular waveguides are connected to each other by a change in the critical fields of the fundamental mode and a hetero-symmetry of the higher-order modes that depends on the specific equilibrium state and precursors the doublet split of azimuthal modes in full tubes. In the remainder of this paper, we shall make this connection between planar waveguides and tubes complete by studying the evolution between them of the full dispersion of the propagating modes with $k\neq 0$. We will see that the (hetero-)symmetry of the higher-order modes is not just some by-product of curvature-induced chirality, but indeed crucially influences the hybridization between the different dispersion branches. Lastly, we will witness the role of the edge modes in planar waveguides on the formation of the nanotube spectrum.

We start in the $\varphi$-saturated regime at $B=\SI{200}{\milli\tesla}$, where the mode parities are well-preserved during the curvature transition. Thus, figure \figref{fig:dispersion_evolution}{a} shows the evolution of the spin-wave dispersion from the transversally-magnetized planar waveguide to the closed vortex-state nanotube. The lines are again colored according to the mode parity. At $\kappa_\mathrm{rel}=0$, the dispersion is again composed of the two lower-frequency edge modes and a couple of bulk modes at higher frequencies, all with parity $P_\nu = \pm 1$ alternating between even and odd \footnote{It can be seen that the edge modes exhibit some hetero-symmetry at large wave vectors. In our calculations, this hetero-symmetry disappeared even for the smallest $\kappa_\mathrm{rel}>0$. We are therefore not sure whether it is supposed to be there, or related to issues during the calculation of the ground state which can appear if the system is too symmetric.}. All modes exhibit positive group velocity $\partial\omega/\partial k$ as they are in the Damon-Eshbach geometry $\bm{k}\perp\bm{m}_0$. With increasing relative curvature $\kappa_\mathrm{rel}$, the dispersion becomes more and more asymmetric in the wave vector $k$, leading to the  nonreciprocal propagation of spin waves along the curved waveguides. This curvature-induced nonreciprocity is of dipolar origin and originates from a geometric contribution to the dynamic volume charges \cite{shekaNonlocalChiralSymmetry2020}. In contrast to the curvature-induced DMI discussed before, this symmetry breaking is of a nonlocal nature and can be described with the increasing toroidal moment $\bm{\tau}\sim\int \bm{r}\times\bm{m}_0\ \mathrm{d}V$ of the equilibrium state \cite{korberCurvilinearSpinwaveDynamics2022}. In fact, dipolar nonreciprocity of spin waves occurs as soon as $\bm{k}\cdot\bm{\tau}\neq 0$, a reflection of the Neumann principle. Nonlocal chiral symmetry breaking has been described for curvilinear shells already in many works \cite{landerosDomainWallMotion2010,otaloraCurvatureInducedAsymmetricSpinWave2016,otaloraAsymmetricSpinwaveDispersion2017,shekaNonlocalChiralSymmetry2020,korberCurvilinearSpinwaveDynamics2022,gallardoHighSpinwaveAsymmetry2022, gallardoUnidirectionalChiralMagnonics2022} and for other cases, for examples in Refs.~\cite{gallardoSpinwaveNonreciprocityMagnetizationgraded2019,henryUnidirectionalSpinwaveChanneling2019}. An elaborate discussion for curvilinear shells is found, for example, in Ref.~\citenum{shekaNonlocalChiralSymmetry2020} or in Sec.~8.3 of Ref.~\citenum{korberSpinWavesCurved2023}. 

Previously reported for the modes in closed nanotubes, it is not surprising that the modes (including the edge modes) in transversally magnetized curved waveguides also exhibit this nonreciprocity. As seen in Fig.~\figref{fig:dispersion_evolution}{a} it can lead to a level-crossing of different modes. However, while modes with different parities exhibit true crossings (seen for $\kappa_\mathrm{rel}=0.75$), modes with the same parity do not cross and exhibit level gaps \cite{trevillian_formation_2024}. For $\kappa_\mathrm{rel}=0.96$ and $\kappa_\mathrm{rel}=0.999$, these gaps formed between all even-parity modes. The anti-crossings appear due to dipole-dipole hybridization allowed between modes of the same parity or, more generally, between modes of the same well-defined irreducible representation. Loosely speaking, this hybridization appears when two modes have a spatial overlap $\langle \bm{\eta}_\nu \vert \bm{\eta}_\mu\rangle \neq 0$, which, in the present case, is only satisfied between equal-parity modes. One can nicely observe how, increasing $\kappa_\mathrm{rel}$ from 0 to 1, the edge modes increase in frequency and merge with the rest of the spectrum via hybridization, thus playing an integral part on the formation of the nanotube spectrum. Again, upon closing the gap from $\kappa_\mathrm{rel}=0.999$ to the full tube with $\kappa_\mathrm{rel}=1$, all odd-parity modes vanish and the even modes become doublets of counter-propagating azimuthal modes.

The same transition is not as well-defined at lower external fields, \textit{e.g.}, at $B=\SI{25}{\milli\tesla}$ in Fig.~\figref{fig:dispersion_evolution}{b}, when we transform from the bulk-tilted planar waveguide into the helical-state tube. At such lower fields, the equilibrium magnetization has a significant component in the $z$-direction. On one hand, this induces the backward-volume character of the propagating modes, with regions of the wave-vector space with negative group velocities. On the other hand, it leads to a parity loss of the higher-order modes with increasing $\kappa_\mathrm{rel}$, as described in Sec.~\ref{sec:parityloss}. One can see in Fig.~\figref{fig:dispersion_evolution}{b} that, with increasing $\kappa_\mathrm{rel}$, dispersion asymmetry still increases, as, even at such small fields, the equilibrium becomes partially oriented in the $\varphi$-direction and acquires a toroidal moment component parallel to the wave vector of the modes. At the same time, the curvature-induced parity loss of the modes destroys their arrangement according to distinct symmetries (into distinct irreducible representations), ultimately leading to a hybridization between all branches. This is clearly seen in Fig.~\figref{fig:dispersion_evolution}{b} as a large number of anti-crossings. 

Finally, at $B=\SI{0}{\milli\tesla}$ in Fig.~\figref{fig:dispersion_evolution}{c}, when transforming the longitudinally magnetized planar waveguide into the axial-state tube, the higher-order modes also become hetero-symmetric. However, lacking a toroidal moment, $\bm{\tau}=0$, there is no emergent nonreciprocity along the $z$-direction. Because of this, different branches do not approach each other and, therefore, hybridization is not observed for this particular geometry.

\section{Conclusions}\label{sec:summary}

In this paper, we provided new insights into curvilinear magnetization dynamics by systematically shedding light on the relation between magnon dynamics in flat and tubular geometries, combining aspects of symmetry, geometry, and topology. Using a finite-element dynamic-matrix method allowed for a comprehensive study of the smooth transition between both geometries across a wide range of magnetization states. By analyzing the dependence of the fundamental magnon mode on an applied transversal field, we traced the evolution of the critical saturation fields involved when rolling up the waveguide cross-section. During this geometric transition, the transversal saturation field strongly decreases due to the enhanced closure of the magnetic flux outside the sample. However, instead of vanishing, the saturation field is increasingly dominated by the curvature-induced anisotropy field, which originates from the exchange interaction and is well-known in magnetic nanotubes and other systems. 

After this initial characterization, we extensively studied the mirror symmetry (parity) of the higher-order magnon modes,  a key determinant of their dynamic properties. In the flat case, all modes are either fully mirror symmetric (even) or anti-symmetric (odd). However, this parity is lost with increasing curvature due to a chiral contribution to the exchange interaction that emerges in the curvilinear frame of reference. It is fundamentally distinct from the dipole-induced hetero-symmetry of Damon-Eshbach waves in thin films and only occurs when the waveguides are still partially magnetized along their axis and not fully transversally saturated. Therefore, this parity loss is a precursor of the topological magnon Berry phase that appears when the cross-section is closed to a tube and splits the degeneracy of counter-propagating azimuthal modes. Above the transversal saturation field, the chiral contribution to the exchange interaction does not contribute to the magnon spectrum, preserving parity of the higher-order modes for all curvatures. 

Parity preservation has crucial implications for the linear coupling between the different modes. Indeed, above saturation, only modes of the same parity hybridize, leading to anti-crossings in the dispersion of the modes propagating with nonzero wavevector along the waveguide axes. In this context, we shed light on the hybridization of the edge and bulk modes of transversally magnetized rectangular waveguides when rolling them up, attesting to the edge modes' integral role in forming the nanotube spectrum. On the other hand, below saturation, where the parity is not preserved, all propagating modes can hybridize, leading to a more complex and rich spectrum. We note that this parity loss will affect not only the hybridization of the modes but also their susceptibility to high-frequency external fields, which is also greatly determined by the symmetry of modes. We expect these predictions to hold not only for tubular geometries but also for other curvilinear waveguides with bound cross-sections, such as crescent-shaped nanorods, nanotrenches, or even tube segments with polygonal cross-sections.

\begin{acknowledgements}
We are thankful to Vadym Iurchuk and Yves Henry for fruitful discussions. Financial support by the Deutsche Forschungsgemeinschaft (DFG) within the program KA 5069/3-1 is gratefully acknowledged.
\end{acknowledgements}



%

\end{document}